# The Highest Expected Reward Decoding for HMMs with Application to Recombination Detection

Michal Nánási, Tomáš Vinař, and Broňa Brejová

Faculty of Mathematics, Physics, and Informatics, Comenius University,
Mlynská Dolina, 842 48 Bratislava, Slovakia

**Abstract.** Hidden Markov models are traditionally decoded by the Viterbi algorithm which finds the highest probability state path in the model. In recent years, several limitations of the Viterbi decoding have been demonstrated, and new algorithms have been developed to address them (Kall et al., 2005; Brejova et al., 2007; Gross et al., 2007; Brown and Truszkowski, 2010).
In this paper, we propose a new efficient highest expected reward decoding algorithm (HERD) that allows for uncertainty in boundaries of individual sequence features. We demonstrate usefulness of our approach on jumping HMMs for recombination detection in viral genomes.

## 1 Introduction

Hidden Markov models (HMMs) are an important tool for modeling and annotation of biological sequences and other data, such as natural language texts. The goal of sequence annotation is to label each symbol of the input sequence according to its meaning or a function. For example, in gene finding, we seek to distinguish regions of DNA that encode proteins from non-coding sequence. An HMM defines a probability distribution $\Pr(A|X)$ over all annotations $A$ of sequence $X$. Typically, one uses the well-known Viterbi algorithm (Forney Jr., 1973) or its variants for more complex models (Brejova et al., 2007) to find the annotation with the highest overall probability $\arg\max_A Pr(A|X)$. In this paper, we design an efficient HMM decoding algorithm that finds the optimal annotation for a different optimization criterion that is more appropriate in many applications.

In recent years, several annotation strategies were shown to achieve better performance than the Viterbi decoding in particular applications (Kall et al., 2005; Gross et al., 2007; Brown and Truszkowski, 2010). Generally, they can be expressed in the terminology of *gain functions* introduced in the context of stochastic context-free grammars (Hamada et al., 2009). In particular, we choose a gain function $G(A, A')$ which characterizes similarity between a proposed annotation $A$ and the (unknown) correct annotation $A'$. The goal is then to find the annotation $A$ with the highest expected value of $G(A, A')$ over the distribution of $A'$ defined by the HMM conditioning on sequence $X$. That is, we maximize $E_{A'|X}[G(A, A')] = \sum_{A'} G(A, A')P(A'|X)$.

In this framework, the Viterbi decoding optimizes the identity gain function $G(A, A') = [A = A']$, that is the gain is 1 if we predict the whole annotation exactly correctly and 0 otherwise. There may be many high-probability annotations besides the optimal one, and they are disregarded by this gain function, even though their consensus may suggest a different answer that is perhaps more accurate locally. On the other hand, the posterior decoding (Durbin et al., 1998) predicts at each position a label that has the highest posterior probability at that position, marginalizing over all annotations. Therefore it optimizes the expected gain under the gain function that counts the number of correctly predicted labels in $A$ with respect to $A'$.

These two gain functions are extremes: the Viterbi decoding assigns a positive gain to the annotation only if it is completely correct, while the posterior decoding gain function rewards every correct label. It is often appropriate to consider gain functions in between these two extremes. For example, in the context of gene finding, Gross et al. (2007) use a gain function that assigns a score $+1$ for each correctly predicted coding region boundary and score $-\gamma$ for predicted boundary that is a false positive. Indeed, one of the main objectives of gene finding is to find exact positions of these boundaries, since even a small error may change the predicted protein significantly. Parameter $\gamma$ in the gain function controls the trade-off between sensitivity and specificity.

While the coding region boundaries are well defined in gene finding and it is desirable to locate them precisely, in other applications, such as transmembrane protein topology prediction, we only wish to infer the approximate locations of feature boundaries. The main reason is that the underlying HMMs do not contain enough information to locate the boundaries exactly, and there are typically many annotations of similar probability with slightly different boundaries. This issue was recently examined by Brown and Truszkowski (2010) in a Viterbi-like setting, where we assign gain to an annotation, if all feature boundaries in $A$ are within some distance $W$ from the corresponding boundary in the correct annotation $A'$. Unfortunately, the problem has to be addressed by heuristics, since it is NP-hard even for $W = 0$.

In this paper, we propose a new gain function in which each feature boundary in $A$ gets score $+1$ if it is within distance $W$ from the corresponding boundary in $A'$, and score $-\gamma$ otherwise. Our definition allows to consider nearby boundary positions as equivalent as in Brown and Truszkowski (2010), yet it avoids the requirement that the whole annotation needs to be essentially correct to receive any gain at all. Another benefit is that our gain function can be efficiently optimized in time linear in the length of the input sequence.

We apply our algorithm to the problem of detecting recombination in the genome of the human immunodeficiency virus (HIV) with jumping HMMs (Schultz et al., 2006). A jumping HMM consists of a profile HMM (Durbin et al., 1998) for each known subtype of HIV. Recombination events are represented by a special jump transitions between different profile HMMs. The goal is to determine for a new HIV genome whether it comes from one of the known subtypes or whether it is a recombination of several subtypes. However, the exact position of a breakpoint can be difficult to determine, particularly if the two recombining strains were very similar near the recombination point. Our gain function therefore corresponds very naturally to this problem. It scores individual predicted recombination points but allows some tolerance in their exact placement.

## 2    HERD: The Highest Expected Reward Decoding

In this section, we propose a new gain function and describe an algorithm for finding the annotation with the highest expected gain. Our algorithm is a non-trivial extension of the maximum expected boundary accuracy decoding (Gross et al., 2007).

*Hidden Markov models and notation.* A *hidden Markov model (HMM)* is a generative probabilistic model with a finite set of states $V$ and transitions $E$. There is a single designated start state $s$ and a final state $t$. The generative process starts in the start state, and in each round it emits a single symbol $x_i$ from the *emission probability* distribution $e_{v_i,x_i}$ of the current state $v_i$, and then changes the state to $v_{i+1}$ according to the *transition probability* distribution $a_{v_i,v_{i+1}}$. The generative process continues until the final state is reached. Thus, the joint probability of generating a sequence $X = x_1, \ldots, x_n$ by a state path $\pi = s, v_1, \ldots, v_n, t$ is $\Pr(\pi, X) = a_{s,v_1} \cdot \prod_{i=1}^{n} e_{v_i,x_i} \cdot a_{v_i,v_{i+1}}$, where $v_{n+1} = t$. In other words, the HMM defines a probability distribution $\Pr(\pi, X)$ over all possible sequences $X$ and state paths $\pi$, or perhaps more appropriately, for a given sequence $X$, the HMM defines a conditional distribution over all state paths $\Pr(\pi \mid X)$.

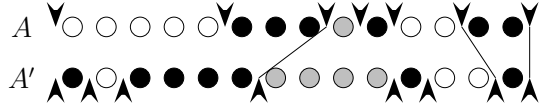

**Fig. 1.** Example of buddy pairs in two annotations over three colors (white, gray, black) for $W = 3$. Boundaries are shown by arrows, buddy pairs are connected by lines. The second boundary in $A$ does not have a buddy pair due to condition (ii), whereas the fourth and fifth boundary due to condition (iii). In this example, $G(A, A') = 3 - 4\gamma$.

Our aim is to produce a *annotation* of an input sequence $X$, i.e. to label each symbol of $X$ by a color corresponding to its function (e.g., coding or non-coding in the case of gene finding, or a virus subtype in case of recombination detection). Position $i$ in the annotation $A = a_1 \ldots a_n$ is a *boundary*, if $a_i$ and $a_{i+1}$ are different colors. For convenience, we consider positions 0 and $n$ as boundaries. A *feature* is a region between two consecutive boundaries.

To use HMMs for sequence annotation, we color each state $v$ by a color $c(v)$. Every state path $\pi = s, v_1, \ldots, v_n, t$ thus implies an annotation $c(\pi) = c(v_1) \ldots c(v_n)$. In general, multiple states can have the same color, and several state paths may produce the same annotation. Thus, HMMs also define a probability distribution over annotations $A$, where $\Pr(A \mid X) = \sum_{\pi: c(\pi) = A} \Pr(\pi \mid X)$.

*The highest expected reward decoding problem.* To formally define our problem, we first define a gain function $G(A, A')$ characterizing similarity between any two annotations $A$ and $A'$ of the same sequence. We assign positive score to a boundary in $A$, if $A'$ contains a corresponding boundary sufficiently close so that they can be considered equivalent. This notion of closeness is formalized in the following definition (see also Figure 1).

**Definition 1.** *Let $A$ and $A'$ be two annotations of the same sequence. Boundaries $i$ in $A$ and $j$ in $A'$ are called* buddies *if (i) both of them separate the same pair of colors $c_1$ and $c_2$, (ii) $|i - j| < W$, and (iii) there is no other boundary at positions $\min\{i, j\}, \ldots, \max\{i, j\}$ in either $A$ or $A'$.*

Condition (iii) enforces that each boundary in $A'$ is a buddy to at most one boundary in $A$. It also ensures that if $i$ and $j$ are buddies, then $j$ cannot be inside any feature of $A$ that is not adjacent to the boundary $i$, avoiding counter-intuitive scores in annotations with short features.

**Definition 2 (Highest expected reward decoding problem).** *Let gain function $G(A, A')$ assign score $+1$ to each boundary in $A$ if it has a buddy in $A'$ and score $-\gamma$ to boundaries in $A$ without a buddy. In the* highest expected reward decoding (HERD)*, we seek the annotation $A$ maximizing the expected gain $E_{A' \mid X}[G(A, A')] = \sum_{A'} G(A, A') \Pr(A' \mid X)$, where the conditional probability $\Pr(A' \mid X)$ is defined by the HMM as $\sum_{\pi: c(\pi) = A'} \Pr(\pi, X) / \Pr(X)$.*

Note that our objective $E_{A' \mid X}[G(A, A')]$ can be further decomposed. In particular, from linearity of expectation, $E_{A' \mid X}[G(A, A')] = \sum_{i \in B(A)} R_\gamma(p_{A,i})$, where $B(A)$ is the set of all boundaries in

**Fig. 2.** Illustration of annotations contributing probability to $p(i, c_1, c_2, w_L, w_R)$ for $W = 3$.

$A$, $p_{A,i}$ is the posterior probability in the HMM that the boundary $i$ in $A$ has a buddy, and $R_\gamma(p) = p - \gamma \cdot (1-p)$ is the expected score (reward) for a boundary with posterior probability $p$.

The HERD algorithm computes posterior probabilities and expected rewards for all possible boundaries and then uses dynamic programming to choose an annotation with the highest possible sum of expected rewards in its boundaries. The details of the algorithm are described below.

*Expected reward of a boundary.* To compute the posterior probability $p_{A,i}$ that a boundary $i$ in $A$ has a buddy in $A'$ sampled from the HMM, it is sufficient to examine only a local neighborhood of boundary $i$ in $A$. In particular, let $c_1$ and $c_2$ be the two colors separated by this boundary and $n_L$ and $n_R$ be the lengths of the two adjacent features. If $n_L \leq W$, the leftmost possible position of the buddy in $A'$ is $i - n_L + 1$, otherwise it is $i - W + 1$; a symmetric condition holds for the rightmost position. Therefore, if $A$ has a buddy in $A'$, it must be in the interval $[i - w_L + 1, i + w_R - 1]$, where $w_L = \min\{W, n_L\}$, and $w_R = \min\{W, n_R\}$. If we denote by $p(i, c_1, c_2, w_L, w_R)$ the sum of probabilities of all annotations $A'$ that have a buddy for boundary $i$ in the interval $[i - w_L + 1, i + w_R - 1]$ (see Figure 2), the expected reward of boundary $i$ will be $R_\gamma(p(i, c_1, c_2, w_L, w_R))$.

Probability $p(i, c_1, c_2, w_L, w_R)$ can be expressed as a sum of simpler terms, one for each possible position $j$ of the buddy in $A'$:

$$p(i, c_1, c_2, w_L, w_R) = \sum_{j=i-w_L+1}^{i} \Pr(a_{j\ldots i+1} = c_1 c_2^{i-j-1} \mid X) + \sum_{j=i+1}^{i+w_R-1} \Pr(a_{i\ldots j+1} = c_1^{j-i-1} c_2 \mid X).$$

Note that if the buddy is at position $j \leq i$, this position needs to have color $c_1$ and all successive positions up to $i+1$ need to have color $c_2$, otherwise there would be a different boundary between $i$ and $j$ in $A'$. However, positions outside of interval $j \ldots, i+1$ can be colored arbitrarily. Similarly for the buddy at position $j > i$, all positions from $i$ up to $j$ need to have color $c_1$ and position $j+1$ color $c_2$. Also note that all terms in the sum represent disjoint sets of annotations, and therefore we are justified to compute the probability of the union of these sets by a sum. All terms of this sum can be computed efficiently, as described at the end of this section.

*Finding the annotation with the highest expected reward.* Once the expected rewards $R_\gamma(p(i, c_1, c_2, w_L, w_R))$ are known for all possible boundaries, we can compute the annotation $A$ with the highest expected gain by dynamic programming. We can view the algorithm as the computation of the highest-weight directed path between two vertices in a directed acyclic graph in which each path corresponds to one annotation and its weight to the expected gain.

In particular, the graph has a vertex $(i, c, w)$ for each position $i$, color $c$ and window length $w \leq W$. This vertex represents a boundary at position $i$ between an unspecified color on the left and the color $c$ on the right, where the adjacent feature of color $c$ has length exactly $w$ if $w < W$, or at least $W$ otherwise. If $w < W$, we will connect vertex $(i, c, w)$ with vertices $(i + w, c', w')$ for all colors $c'$ and lengths $w' \leq W$. Each such edge will have weight $R_\gamma(p(i + w, c, c', w, w'))$, representing the expected reward of boundary at position $i + w$. If $w = W$, we connect vertex $(i, c, w)$ with vertices $(i + w'', c, c', w')$ for all $w'' \geq W$, $w' \leq W$ and color $c'$ by *long-distance edges*. The weight of such edges will be $R_\gamma(p(i + w'', c, c', W, w'))$.

To finish the construction, we will assume that positions $0$ and $n + 1$ are labeled by special colors $c_s$ and $c_f$ and that these two features have corresponding nodes in the graph. We also add a starting vertex $(-1, c_s, 1)$ and connect it to vertices $(0, c, w)$ according to normal rules. The annotation with the highest reward corresponds to the highest-weight path from vertex $(-1, c_s, 1)$ to vertex $(n, c_f, 1)$.

In this construction, the number of long-distance edges is quadratic in the length of sequence $X$, leading to an inefficient algorithm. Fortunately, the cost of a long-distance edge from $(i, c, w)$ to $(i + w'', c, c', w')$ does not depend on index $i$, only on $i + w''$. Therefore, every long-distance edge can be replaced by a path through a series of special collector vertices of the form $(i, c)$ for a position $i$ and color $c$. There is an edge of weight $0$ from $(i, c, W)$ to $(i + W, c)$ for entering the collector path at an appropriate minimum distance from $i$, edge of weight $0$ from $(i, c)$ to $(i + 1, c)$ for continuing in the collector path, and an edge of weight $R_\gamma(p(i, c, c', W, w'))$ for leaving the collector path from vertex $(i, c)$ to vertex $(i, c', w')$. This modified graph has $O(nWC)$ vertices and $O(nW^2C^2)$ edges, where $n$ is the length of the sequence, $W$ is the size of the window and $C$ is the number of different colors in the HMM.

*Implementation details and running time.* The only remaining detail is the computation of the posterior probabilities of the form $\Pr(a_{i...i+w} = c_1 c_2{}^w \mid X)$ and $\Pr(a_{i...i+w} = c_1^w c_2 \mid X)$ needed to compute $p(i, c, c', w, w')$. We will show how to compute the first of these two quantities, the second is analogous.

First, we use the standard forward algorithm to compute $F[i, v]$ which is the sum of probabilities of all state paths ending in state $v$ after generating the first $i$ symbols from $X$. We use a modified backward algorithm to compute $B[i, v, w]$, which is the sum of probabilities of all state paths generating symbols $x_i \ldots x_n$ that start in state $v$ and generate the first $w$ symbols in the states of color $c(v)$. Values $B[i, v, 1]$ are computed by the standard backward algorithm and $B[i, v, w]$ for $1 < w \leq W$ is computed by the following simple formula:

$$B[i, v, w] = \sum_{v \to v', c(v) = c(v')} B[i + 1, v, w - 1] \cdot e_{v, x_i} \cdot a_{v, v'}.$$

Finally, the desired posterior probability is obtained by combining forward and backward probabilities over all transitions passing from color $c_1$ to color $c_2$ at position $i$:

$$\Pr(a_{i...i+w} = c_1 c_2{}^w \mid X) = \sum_{v \to v', c(v) = c_1, c(v') = c_2} F[i, v] \cdot a_{v, v'} \cdot B[i + 1, v', w].$$

The standard forward algorithm works in $O(n|E|)$ time, our extended backward algorithm takes $O(nW|E|)$ time. Posterior probabilities are summarized from these quantities also in $O(nW|E|)$ time. Finally, we construct and search the graph in $(nW^2C^2)$ time. Thus the overall running time is $O(nW|E| + nW^2C^2)$. Note that the time is linear in the sequence length, which is very important for applications in genomics, where we analyze very long genomic sequences.

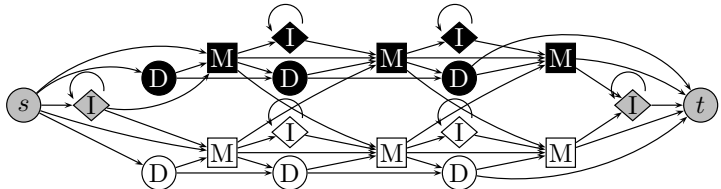

**Fig. 3.** A small example of a jumping HMM with two profile HMMs. For readability, jumping transitions between match states (M) and insert (I) or delete (D) states are not shown.

## 3  Application to Viral Recombination Detection

Most HIV infections are caused by HIV-1 group M viruses. These viruses can be classified by a phylogenetic analysis into several subtypes and sub-subtypes. However, some HIV genomes are a mosaic of sequences from different subtypes resulting from recombination between different strains (Robertson et al., 2000). Our goal is to classify whether a newly sequenced HIV genome comes entirely from one of the known subtypes or whether it is a recombination of different subtypes, which is important for monitoring the HIV epidemics.

Schultz et al. (2006) propose to detect recombination by jumping HMMs. In this framework, multiple sequence alignment of known HIV genomes is divided into parts corresponding to individual subtypes or sub-subtypes, and a profile HMM is built for each. A profile HMM (Durbin et al., 1998) represents one column of alignment by a match state, insert state and delete state. Emission probabilities of the match state correspond to frequencies of symbols in that alignment column. Insert state represents sequences inserted immediately after the column and delete state is a silent state allowing to bypass the match state without emitting any symbols, thus corresponding to a deletion. A jumping HMM also contains low probability jump transitions between profile HMMs corresponding to individual subtypes, as shown in Figure 3.

To use a jumping HMM for recombination detection, we color each state by its subtype. Then, boundaries in annotation correspond to recombination breakpoints. Schultz et al. (2006) use the Viterbi algorithm and report the annotation corresponding to the most probable state path. However, the same annotation can be obtained by many different state paths corresponding to different alignments of the input sequence to the profile HMMs. Even though in the latest version of their software (Schultz et al., 2009) they augment the output by display of posterior probabilities, they still output only a single annotation obtained by the Viterbi algorithm. Since we are not interested in the alignment, only in the annotation, it is more appropriate to use the most probable annotation instead of the most probable path. However, the problem of finding the most probable annotation is NP-hard for many HMMs (Brejova et al., 2007), and jumping HMMs due to their complicated structure with many transitions between states of different color are likely to belong to this class.

The HERD bypasses this computational difficulty by maximizing a different gain function that scores individual breakpoints rather than the whole annotation. Compared to the Viterbi algorithm, our algorithm considers all possible state paths (alignments) contributing to the resulting annotation. In addition, our algorithm considers nearby potential recombination points as equivalent, since in practice it is difficult to determine the exact recombination point, particularly in strongly conserved regions or between related subtypes.

The use of jumping HMMs on HIV genomes is relatively time consuming, as a typical HIV genome has length almost 10,000 bases and the jumping HMM has 7,356,740 transitions. Schultz et al. (2006) use the beam search heuristic to speed up the Viterbi algorithm. Unfortunately, this heuristic is not applicable in our case, and our algorithm is also asymptotically slower than the Viterbi algorithm by a factor of $W$. To reduce the running time, we use a simple anchoring strategy, similar to heuristics frequently used in global sequence alignment (Kurtz et al., 2004). We choose as anchors 19 well-conserved portions of the HIV multiple alignment and align the consensus sequence of each anchor to the query sequence. In the forward and backward algorithm we then constrain the alignment of the query to the profile HMMs so that the position of the anchor in the query aligns to its known position in the profile HMM. We also extend the algorithm described above to handle silent states by modifying the preprocessing stage.

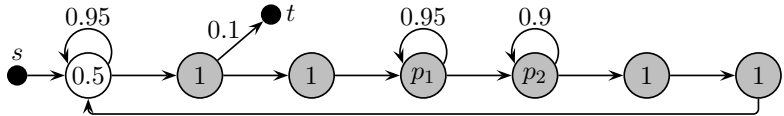

**Fig. 4.** A toy HMM emitting symbols over the binary alphabet, where the numbers inside states represent the emission probability of the symbol 1. States $s$ and $t$ are silent. The HMM outputs alternating white regions of mean length 20 and gray regions of mean length 34. The distribution of symbols is constant in the white regions, while in the gray regions, it changes towards their right ends. The gray regions are bounded by a two-symbol signal 11 on both sides. The HMM was inspired by models of CT-rich intron tails in gene finding (Brejova et al., 2007).

## 4 Experiments

*Toy sequence annotation HMM.* We have first tested our algorithm on data generated from a toy HMM in Figure 4. This HMM has multiple state paths for a given annotation, and we have previously demonstrated that the most probable annotation is more accurate than the annotation corresponding to the most probable state path found by the Viterbi algorithm (Brejova et al., 2007).

Table 1 shows different measures of accuracy for several decoding algorithms on 5000 randomly generated sequences of mean length about 500. We report results for two sets of parameter values of the model, however, the trends observed in the table generally hold also for other combinations of $p_1$ and $p_2$. As we have shown earlier, the extended Viterbi algorithm (EVA) (Brejova et al., 2007) for finding the most probable annotation generally outperforms the Viterbi algorithm. The HERD with parameters $W = 5$ and $\gamma = 1$ is more accurate when the performance is measured by its own gain function, which is not surprising, since the data and baseline predictions are generated from the same model as is used for annotation. On the other hand, the HERD colors fewer bases correctly and tends to place boundaries on average further away from the correct ones than the EVA. This is also not unexpected, as the HERD explicitly disregards small differences in the boundary position. We have also measured sensitivity and specificity in predicting individual features. Here the HERD works better than the EVA for some parameter settings (e.g. $p_1 = p_2 = 0.9$ in the table), but not for others. We have also run the HERD with $W = 1$, which is equivalent to maximum expected boundary accuracy decoding (Gross et al., 2007). The accuracy of this decoding is very poor for $\gamma = 1$, but markedly improves for lower penalty $\gamma = 0.1$. The reason is that for $W = 1$, we sum over fewer state paths and therefore the posterior probability of a boundary rarely reaches the threshold $1/2$ necessary to achieve positive expected reward at $\gamma = 1$.

*HIV recombination detection.* Table 2 shows the accuracy of the HERD on predicting recombination in HIV genomes. In all tests, we have used the sequence data and the jumping of Schultz et al.

**Table 1.** The accuracy on synthetic data generated from the HMM in Figure 4. (i) Fraction of the bases colored by the same color by the algorithm and the correct annotation (baseline). (ii) Gain function $G(A, A')$ between the prediction and the baseline for $W = 5$ and $\gamma = 1$. (iii) A feature is predicted correctly if there is a corresponding feature of the same color in the baseline with both boundaries in distance less than 5. Specificity (sp.) is the fraction of all predicted features that are correct, and sensitivity (sn.) is the fraction of baseline features that are correctly predicted. (iv) Mean distance between the baseline and predicted boundary for all correctly predicted features.

| Algorithm | % bases correct[(i)] | Gain [(ii)] | Feature sp.[(iii)] | Feature sn.[(iii)] | Avg. dist. |
|---|---|---|---|---|---|
| **HMM parameters $p_1 = 0.9$, $p_2 = 0.9$** | | | | | |
| HERD $W = 5, \gamma = 1$ | 85.6% | 12.0 | 71.0% | 58.0% | 1.9 |
| HERD $W = 1, \gamma = 1$ | 47.5% | 3.0 | 55.1% | 17.8% | 0.0 |
| HERD $W = 1, \gamma = 0.1$ | 90.4% | 2.4 | 51.8% | 66.0% | 0.9 |
| Viterbi | 89.4% | 8.9 | 66.3% | 47.3% | 0.7 |
| Extended Viterbi | 91.2% | 10.3 | 69.9% | 56.2% | 0.8 |
| **HMM parameters $p_1 = 0.7$, $p_2 = 0.8$** | | | | | |
| HERD $W = 5, \gamma = 1$ | 74.3% | 5.0 | 47.4% | 30.9% | 1.7 |
| HERD $W = 1, \gamma = 1$ | 47.5% | 3.0 | 55.0% | 17.7% | 0.0 |
| HERD $W = 1, \gamma = 0.1$ | 79.6% | -2.7 | 38.2% | 43.9% | 0.9 |
| Viterbi | 75.0% | 3.6 | 51.2% | 25.7% | 0.4 |
| Extended Viterbi | 79.7% | 4.1 | 49.0% | 31.3% | 0.6 |

(2006), though in most tests we have increased the jump probability $P_j$ from $10^{-9}$ to $10^{-5}$. With the original value, the HERD rarely predicts any recombination, since the posterior probability of a breakpoint has to be at least $1/2$ for $\gamma = 1$ to receive a positive score, and with the lower jumping probability we usually do not reach such a level of confidence. We have conducted the tests on a 1696 column region of the whole genomic alignment, starting at position 6925. This restriction allowed us to test higher number of sequences than Schultz et al. (2006) reasonably fast.

The first set of tests was done on 62 real HIV sequences without known recombination. These sequences were selected from subtypes A1, B, C, D, F1 (10 sequences from each subtype) and G, A2, F2 (5, 3, and 4 sequences respectively) and omitted from the training set (except for subtypes A2, F1 and F2 which have very few samples). As we can see in Table 2, the Viterbi algorithm always predicts the correct result. Our algorithm on the jumping HMM with the original low jumping probability $P_j = 10^{-9}$ also produces correct answer every time. However, the value of $P_j = 10^{-5}$ leads to spurious recombinations predicted in 11.3% of sequences, thus lowering the accuracy.

The second set of sequences contains artificial recombinants. Each of them was created as a combination of two sequences from two different subtypes by alternating regions of length 300. The set contains recombinants between subtype pairs A-B, A-C, A-G, B-C, B-G and C-G, 50 sequences from each pair. Similarly as in the toy HMM case, our algorithm performs worse with respect to the total number of correctly labeled bases and average distance to the correct boundary, but it finds individual features (recombinant regions) with much greater sensitivity and specificity if we allow some tolerance in the boundary placement. For $W = 1$ the HERD has very low accuracy even for lowered penalty $\gamma = 0.1$. This suggests that our generalization of the maximum expected boundary accuracy decoding to $W > 1$ is crucial in this setting.

In the third test, we have used the same procedure as in the previous test to create 170 artificial recombinants between sequences of two sub-subtypes of the same subtype (A1 and A2, F1 and F2) or from two subtypes at a small phylogenetic distance more typical for sub-subtypes (B and D).

9**Table 2.** The accuracy on HIV recombination data. The meaning of the columns is the same as in Table 1, except that we use $W = 10$ and $\gamma = 1$ in the definition of the gain function and correctly predicted features.

| Algorithm | % bases correct[(i)] | Gain [(ii)] | Feature sp.[(iii)] | Feature sn.[(iii)] | Avg. dist. |
|---|---|---|---|---|---|
| **Sequences without recombination** | | | | | |
| HERD, $W = 10, \gamma = 1, P_j = 10^{-9}$ | 100.0% | 2.0 | 100.0% | 100.0% | 0.0 |
| HERD, $W = 10, \gamma = 1, P_j = 10^{-5}$ | 93.7% | 1.5 | 83.9% | 83.9% | 0.0 |
| Viterbi | 100.0% | 2.0 | 100.0% | 100.0% | 0.0 |
| **Sequences with artificial inter-subtype recombination** | | | | | |
| HERD $W = 10, \gamma = 1, P_j = 10^{-5}$ | 92.4% | 2.5 | 62.5% | 56.0% | 2.4 |
| HERD $W = 1, \gamma = 0.1, P_j = 10^{-5}$ | 78.3% | 0.9 | 36.3% | 27.8% | 1.3 |
| Viterbi | 95.1% | 2.0 | 53.1% | 47.1% | 1.8 |
| **Sequences with artificial intra-subtype recombination** | | | | | |
| HERD $W = 10, \gamma = 1, P_j = 10^{-5}$ | 89.6% | 1.6 | 44.6% | 40.7% | 2.7 |
| Viterbi | 88.0% | 1.3 | 32.8% | 26.1% | 2.7 |

The overall accuracy is lower in this test, because it is more difficult to distinguish recombination among more closely related sequences. The HERD is still much more accurate at the feature level and even becomes more accurate than the Viterbi algorithm on the base level.

One issue with our tests is that we have used a lower jump probability $P_j = 10^{-9}$ for sequences without recombination and a higher value $P_j = 10^{-5}$ for sequences with recombination. This distinction is justified by the fact that although recombinant sequences are generally rare, suggesting a low jumping probability, they usually have several recombination points, whose detection then requires a higher value of $P_j$. In practice, when faced with a sequence of unknown origin we propose to first test whether the sequence is likely to be a recombinant, perhaps by a likelihood ratio test with nested models (Felsenstein, 2004) in which $P_j$ is optimized for the input sequence in one model and set to 0 for the null model. If the sequence appears to contain recombination, we can then apply the HERD with the higher value of $P_j$ to determine the breakpoints.

We have also run our algorithm on 12 naturally occurring recombinants, using $W = 10$, $\gamma = 1.5$, and $P_j = 10^{-5}$. Here, we have used the whole length of the sequence. Due to the small number of sequences and uncertain annotation, we do not report the accuracy statistics. Nonetheless, on six sequences the HERD found the correct set of recombining subtypes (on annotated regions); two of them the HERD annotated better than Viterbi (CRF08, CRF12). On the remaining six, the HERD predicted at least one erroneous subtype and often misplaced breakpoints or jumped frequently, but the Viterbi algorithm also made numerous mistakes on two of these sequences.

## 5  Conclusion

In this paper, we have introduced a novel decoding algorithm for hidden Markov models seeking an annotation of the sequence in which boundaries of individual sequence features are at least approximately correct. This decoding is particularly appropriate in situations where the exact boundaries are difficult to determine, and perhaps their knowledge is not even necessary.

We apply our algorithm to the problem of recombination detection in HIV genomes. Here, the Viterbi decoding considers for a given annotation only a single alignment of the query to the profile HMMs and only one placement of breakpoints. In contrast, we marginalize the probabilities over all possible alignments and over nearby placements of recombination boundaries. As a result, we are able to predict individual recombinant regions with greater sensitivity and specificity.

Our experiments also suggest venues for future improvement. First of all, the accuracy results vary with the choice of parameters $P_j$, $W$, and $\gamma$. It remains an open question how to choose these parameters in a principled way. We have also observed that our algorithm does not perform as well as the Viterbi algorithm in finding the exact boundaries. Perhaps this could be solved by a gain function in which a boundary with a more distant buddy gets a smaller score. Similarly, our algorithm performs slightly worse in terms of base-level accuracy, and this shortcoming perhaps could be addressed by adding a positive score for every correctly colored nucleotide to the gain function. In general, the framework of maximum expected gain decoding is very promising, because it allows to tailor decoding algorithm for a specific application domain.

*Acknowledgements.* We would like to thank Dan Brown and Jakub Truszkowski for helpful discussion on related problems. Research of TV and BB is funded by European Community FP7 grants IRG-224885 and IRG-231025.